\documentclass[review]{elsarticle}
\UseRawInputEncoding
\usepackage[colorlinks]{hyperref}
\usepackage{lineno,hyperref}
\usepackage{amsmath}
\usepackage{multirow}
\usepackage{color}

\newcommand{\rev}{\textcolor{black}}

\nolinenumbers

\journal{Journal of theoretical biology}

\begin{document}

\begin{frontmatter}

\title{Broad cross-reactivity of the T-cell repertoire achieves specific and sufficiently rapid target searching}


\author[mymainaddress,mysecondaryaddress]{Jin Xu}

\author[mymainaddress,mysecondaryaddress,mythirdaddress,myfourthaddress]{Junghyo Jo\corref{mycorrespondingauthor}}
\cortext[mycorrespondingauthor]{Corresponding author}
\ead{jojunghyo@kmu.ac.kr.}

\address[mymainaddress]{Asia Pacific Center for Theoretical Physics (APCTP), 67 Cheongam-ro, Pohang 37673, South Korea}
\address[mysecondaryaddress]{Department of Physics, Pohang University of Science and Technology (POSTECH), 77 Cheongam-ro, Pohang 37673, South Korea}
\address[mythirdaddress]{School of Computational Sciences, Korea Institute for Advanced Study (KIAS), 85 Hoegiro, Seoul 02455, South Korea}
\address[myfourthaddress]{Department of Statistics, Keimyung University, 1095 Dalgubeol-daero, Daegu 42601, South Korea}

\begin{abstract}
The molecular recognition of T-cell receptors is the hallmark of the adaptive immunity.
Given the finiteness of the T-cell repertoire, individual T-cell receptors are necessary to be cross-reactive to multiple antigenic peptides.
In this study, we quantify the variability of the cross-reactivity by using a string model that estimates the binding affinity between two sequences of amino acids. We examine sequences of 10,000 human T-cell receptors and 10,000 antigenic peptides, and obtain a full spectrum of cross-reactivity of the receptor-peptide binding.
Then, we find that the cross-reactivity spectrum is broad. Some T cells are reactive to 1,000 peptides, but some T cells are reactive to only one or two peptides.
Since the degree of cross-reactivity has a correlation with the (un)binding affinity of receptors, we further investigate how the broad cross-reactivity affects the target searching of T cells.
High cross-reactive T cells may not require many trials for searching correct targets, but they may spend long time to unbind from incorrect targets. In contrast, low cross-reactive T cells may not spend long time to ignore incorrect targets, but they require many trials for screening correct targets. We evaluate this hypothesis, and show that the broad cross-reactivity of the natural T-cell repertoire can balance the trade-off between the rapid screening and unbinding penalty.
\end{abstract}

\begin{keyword}
adaptive immunity\sep molecular recognition\sep cross-reactivity \sep target searching
\end{keyword}

\end{frontmatter}


\section{Introduction}

T cells are major components of the adaptive immune system. To selectively recognize pathogens, T cells express distinct receptors. Antigen presenting cells engulf pathogens, degrade the pathogens into short peptides, and display these peptides in conjunction with major histocompatibility complexes (pMHC) on their surfaces~\cite{Owen2013}. Subsequently, T cells recognize the processed peptides using T-cell receptors (TCRs). Therefore, the molecular interaction between TCRs and pMHC is crucial for immunological recognition.
The degree of specificity among T cells for pMHCs \rev{has} considerable cross-reactivity~\cite{Mason1998}.  Human have approximately $5 \times 10^6$ distinct T cells~\cite{Warren2011}, whereas unlimited numbers of antigenic peptides exist in principle.
Considering 15-mer peptides, they can have a maximum of $3 \times 10^{19}$ ($\sim 20^{15}$) different combinations of amino acid sequences. The human body is estimated to have $3 \times 10^{13}$ total cells~\cite{Sender2016}. Thus, \rev{finite diversity of T cells is not affordable} to deal with all the peptides specifically due to these physical limitations. In other words, each T cell must recognize multiple antigenic peptides~\cite{Mason1998}. Indeed, each T cell can productively interact with several peptides~\cite{Unanue1984}, which is called T-cell cross-reactivity. Mason emphasized that cross-reactivity is an intrinsic and necessary characteristic of T-cell recognition, and estimated that an individual T cell can respond to $10^6 -10^7$ nonamer peptides, and $10^8$ or more 11-mer peptides~\cite{Mason1998}. Furthermore, a specific peptide can be recognized by different T cells~\cite{Zarnitsyna2013}, which is called peptide cross-reactivity.

For the theoretical estimation of the T-cell cross-reactivity or peptide cross-reactivity, one presumed that every T cell or peptide has uniform degrees of cross-reactivity.
In fact, current experiments are limited to probe the variability of cross-reactivity, because they have mostly examined a few TCRs with a few peptides, and sometimes a few TCRs with a few thousands of peptides~\cite{Birnbaum2014}.
Structure-based approaches are also limited to answer this question because the Protein Data Bank has only about several hundred 3-dimensional structures of TCR-pMHC complexes.
However, we note that recent high-throughput sequencing technologies provide sequence information of TCRs~\cite{Warren2011,Murugan2012,Zvyagin2014}, and antigenic peptides for many infectious diseases~\cite{Vita2015}.
In addition, theoretical immunologists have developed string models that consider sequences of TCR/peptide alphabets, and define molecular recognition based on the matching score between the aligned sequences of TCRs and peptides. The string model is simple, but it has been used to explore fundamental questions in immunology including self/non-self discrimination~\cite{Percus1993}, thymic selection~\cite{DetoursPerelson1999,Detours1999, Chao2005}, and alloreactivity~\cite{Detours2000}.
The simplicity can allow a large-scale analysis on the cross-reactivity between a large numbers of TCRs and peptides.
In this study, we adopt a string model~\cite{Kosmrlj2008} that considers real amino acid sequences of TCRs and peptides, and examine the variability of the cross-reactivity for T cells and peptides by examining their real sequences~\cite{Murugan2012, Vita2015}.

By examining the binding affinity between $10^4$ human TCRs and $10^4$ antigenic peptides, we find that the cross-reactivity is certainly broad. Some T cells are reactive to 1,000 peptides, but some T cells are reactive only one or two peptides.
Next, since the cross-reactivity has a positive correlation with the binding affinity, we further explore the target searching of T cells via the binding and unbinding events with peptides.
The target searching process has been extensively studied in the context of transcription factors binding/unbinding from DNA~\cite{Berg1987,Hippel1986}. The binding affinity determines the time required for transcription factors to scan specific locations on DNA. Strong binding can delay the scanning process due to slow unbinding, while weak binding can overlook target locations~\cite{Gerland2002,Savir2016}. Similarly, high cross-reactive T cells can recognize more targets, but they are less specific to targets. Their strong binding affinity may be beneficial for stable binding to correct targets but detrimental for unbinding from incorrect targets. In this study, we evaluate this hypothesis, and show that the natural T-cell repertoire can balance the trade-off between rapid screening and the unbinding penalty for target searching.

\section{Methods}

\subsection{Model and data}
TCRs have a specific region, complementary-determining region 3 (CDR3), which is critical for recognizing the short peptides in the cove of MHC molecules. Since CDR3 is short, linear sequence pairing models could approximately describe the interaction between CDR3 and pMHC without considering their three-dimensional protein-protein interactions~\cite{Kosmrlj2008,Kosmrlj2009}. Unlike the initial string models using binary sequences~\cite{Percus1993,Segel2001}, a later string model considers amino acid sequences explicitly~\cite{Kosmrlj2008}.
To examine the cross-reactivity of real T cells in this study, we adopt the later model where the binding energy between a TCR and pMHC is defined as
\rev{
\begin{equation}
\label{eq:bindingE}
E(\vec t,\vec p) = \sum_{\rev{i = 5}}^{\rev{10}} {J({t_i},{p_i})}, 
\end{equation}}
which represents the pairwise interaction between the amino acid sequences of a CDR3 of length $L$, $\vec t = ({t_1},{t_2}, \cdots ,{t_L})$, and a peptide, $\vec p = ({p_1},{p_2}, \cdots ,{p_L})$.
The binding energy $J({t_i},{p_i})$ of a pair of amino acids, $t_i$ and $p_i$, is defined by a knowledge-based potential between amino acids. The statistical potentials for each pair of 20 amino acids are represented by the $20\times20$ Miyazawa-Jernigan matrix \cite{Miyazawa1996}, which has been widely used in protein design and folding simulations~\cite{Li1997}, because the statistical potential can provide a reasonable description for protein-protein interactions.
\rev{For the pairwise interaction, we considered that a partial region (5th to 10th residue) of CDR3 contributes mostly for the immunological recognition~\cite{Stadinski2016}.
Therefore, an effective interaction length can be defined as $L_{\text{eff}}=6$.
In this study, to focus on the specific interaction between CDR3 and peptides, we did not explicitly consider the interaction between the remaining part of the TCR excluding the CDR3 part and the MHC molecule as Ref.~\cite{Tsurui2013}, unlike the original model in Ref.~\cite{Kosmrlj2012}.}
This setup defines the event of an immune response when the pairing energy between $\vec t$ and $\vec p$ is larger than a recognition threshold, $E(\vec t,\vec p) > {E_R}$, where $E_R$ is to be defined.

To examine the cross-reactivity of TCRs with the linear binding model, we used real amino acid sequences of TCRs and antigenic peptides. For the TCR sequences, we used a published data on the DNA sequence of $\beta$-chain CDR3 from T cells~\cite{Murugan2012}, which were obtained from blood samples of nine human subjects \cite{Robins2010,Robins2009}. 
\rev{In the database, we used $10^4$ to $10^5$ productive/in-frame sequences of TCRs per subject, which are filtered out from thymic selection.}
In this study, we considered only distinct TCR sequences and ignored the clone size of the T cells because no correlation was observed between the cross-reactivity and the clone size.
Using the DNA codon table, we decoded the nucleotide sequences of DNA into the corresponding amino acid sequences.
\rev{The CDR3 region is defined from the conserved cysteine around the end of the V segment to the third amino acid to the conserved phenylalanine in the J segment ~\cite{Murugan2012}.}
The sequence lengths ranged from 3 to 17 amino acids, and the most frequent length was 12 amino acids. Our analyzed length distribution is consistent with the reported distribution~\cite{Murugan2012}.
For the peptide sequences, we used an open source, Immune epitope database and analysis resource~\cite{Vita2015}, and obtained $10^4$ peptides in human infectious diseases with MHC restriction. Their sequence lengths ranged from 5 to 45 amino acids, and the most frequent length was 15 amino acids.
Since the pairwise binding energy in Eq.~(\ref{eq:bindingE}) depends on the sequence length $L$, longer sequences should have larger binding energies. Therefore, to rule out the effect of the sequence length, we used only 12-mer TCRs ($\sim 10^4$ sequences per subject) and 15-mer peptides ($\sim 10^4$ sequences), \rev{and did not use TCR or peptide sequences with other lengths.}
Then, a 12-mer TCR and a 15-mer peptide can have four possible binding scenarios for $L=12$ depending on where the first site of the TCR meets on the peptide. 
\rev{We randomly selected one binding scenario to represent the stochastic binding between TCRs and peptides.}
Here we also considered the possibility that the reversed \rev{amino acid} sequences of TCRs meet peptides.
Based on the binding energy $E(\vec{t}, \vec{p})$, we defined the reactivity between TCRs and peptides, and calculated their cross-reactivity and target searching time.
The source code to reproduce our results is available in a repository at Jin Xu's GitHub (https://github.com/SunnyXu), which can be accessed from Jin Xu's personal website (https://sites.google.com/site/jinxuphys2018/code/immune-response).


\subsection{Validity of the string model}
The reactivity of TCRs and peptides is determined by their three-dimensional protein structures and dynamics. Thus many factors such as surface shapes of TCRs, loops of CDRs, and a few hot spots of amino acid residues contribute to determine the reactivity~\cite{Sewell2012, Park2013, Chen2018}.
The string model based on one-dimensional amino acid sequences may be too naive to capture such complex immunological recognition.
Nevertheless, if the string model is a reasonable approximation, the simple model has a very strong merit for large-scale screening with cheap computation.
Furthermore, the model is relevant to probe the role of amino acid composition and affinity of TCRs for determining the immunological recognition.

It is challenging to experimentally observe the reactivity between many TCRs and many peptides.
One experimental study reported 2,329 and 4,818 peptide sequences ($L=14$) that induced immune responses for two human TCR CDR3 $\beta$ chains ($L=16$) of Ob1A12 and Ob2F3~\cite{Birnbaum2014}.
Since the amino acid sequences for Ob1A12 and Ob2F3 are available, we examined if the string model predicts high binding energies between the CDR3 and reported peptides.
To obtain control peptides, we generated synthetic peptides of which sequence compositions follow the amino acid frequency distribution of human (infectious, allergic and auto immune) diseases peptides in the IEDB peptide database~\cite{Vita2015}.
Given $\vec{t}$ of Ob1A12 and Ob2F3, we computed $E(\vec{t}, \vec{p})$ for the reported and synthetic peptides $\vec{p}$.
As the recognition threshold $E_R$ decreases, both peptides are more recognized by the two TCRs (Table~\ref{table:model}). 
However, we confirmed that the two TCRs could recognize the reported peptides more successfully than the synthetic peptides.
Although the prediction of the string model depends on the threshold $E_R$, the specificity of Ob1A12 and Ob2F3 was demonstrated by the stronger binding with the reported peptides than the randomly generated peptides.
If one defines a mean binding energy of each amino acid $a_i$ as $\langle E_i \rangle = 1/20 \sum_{j=1}^{20} J(a_i, a_j)$, 20 amino acids can be sorted from the strongest one to the weakest one.
Then, given an amino acid sequence of a peptide, the total mean binding energy can be calculated by using $\langle E_i \rangle$.
We confirmed that the reported peptides ($66.2k_BT$ for 2,329 peptides and $66.7 k_BT$ for 4,818 peptides) had larger total mean binding energy than the synthetic peptides ($63.8 k_BT$) on average, where ${k_B}$ is the Boltzmann constant and $T$ is temperature. Note that at normal body temperature, the thermal energy is ${k_B}T=0.6$ kcal/mol.

Here, we emphasize that the string model is suitable to obtain statistical conclusions from a large-scale analysis.
It is not our aim to predict the exact reactivity between an individual TCR and a specific peptide by using the simple string model.
In this study, we explore how the composition and affinity of amino acids in TCRs affect the cross-reactivity and target searching of TCRs by examining many TCRs and many peptides. Such a large-scale investigation is experimentally infeasible at the moment.
Note that we focus on CDR3 $\beta$ chains although CDR3 $\alpha$ chains also significantly contribute to the immunological recognition~\cite{Hong1992}, because our statistical conclusion can be similarly applicable to CDR3 $\alpha$ chains.

\begin{table}[h]
\centering
\begin{tabular}{ccccc}
\hline
\multirow{2}{*}{$E_R (k_BT)$}  & \multicolumn{2}{c}{Ob1A12} & \multicolumn{2}{c}{Ob2F3}  \\
\cline{2-5}
& Reported ($\%$)& Synthetic ($\%$) & Reported ($\%$) & Synthetic ($\%$)\\
\hline
$15$ & $100$            & $84.1 \pm 0.6$  & $99.9$           & $78.8 \pm 0.9$ \\
$18$   & $84.6 \pm 0.8$   & $43.0 \pm 1.3$  & $75.8 \pm 0.3$   & $34.9 \pm 0.8$ \\
$21$ & $18.6 \pm 0.8$   & $10.4 \pm 0.4$  & $13.4 \pm 0.5$   & $6.6  \pm 0.5$ \\
\hline
\end{tabular}
\caption{Model validation.
Success recognition percentile ($\%$) of peptides by two CDR3 $\beta$ chains of T-cell receptors: Ob1A12 and Ob2F3.
Two groups of peptides are considered: (i) Reported peptides that are known to react with Ob1A12 (2,329 peptides) and Ob2F3 (4,818 peptides) in Ref.~\cite{Birnbaum2014}; (ii) Synthetic peptides that are randomly generated by using the frequency distribution of amino acids for human antigenic peptides in Immune epitope database~\cite{Vita2015}. The success of reactivity between the TCRs and the peptides depends on the recognition threshold $E_R$. The errors are estimated from ten ensembles of synthetically generated peptides, and random binding alignment between the two sequences of TCRs (sequence length $L=16$) and peptides ($L=14$).}
\label{table:model}
\end{table}


\section{Results}

\begin{figure}[h!]
\centering\includegraphics[width=\linewidth]{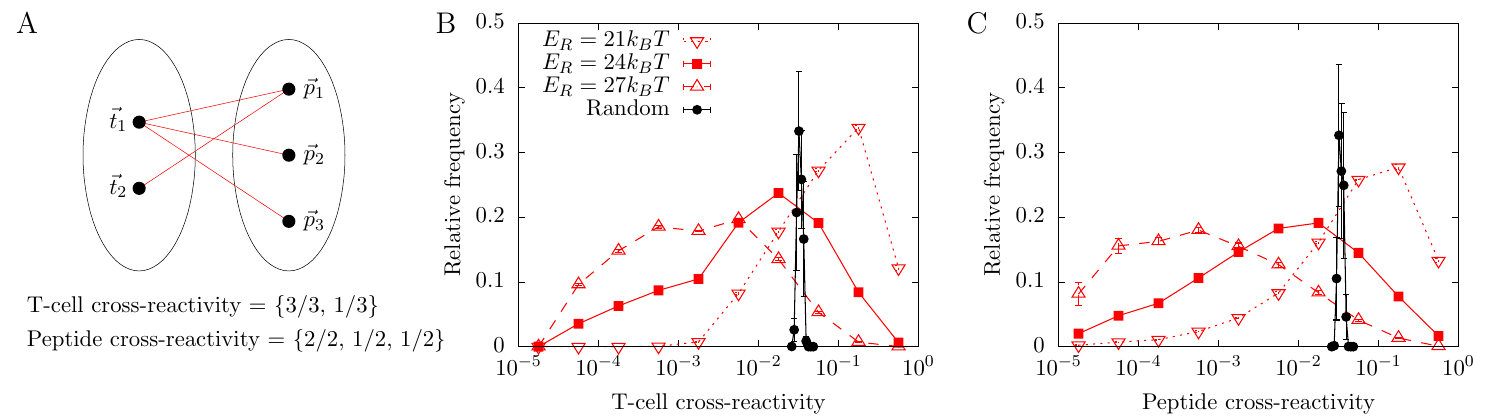}
\caption{T-cell and peptide cross-reactivity. (A) Recognition (red lines) of T-cell receptors (TCR) for antigenic peptides is defined when the pairwise binding energy between a TCR $(\vec t)$ and a peptide $(\vec p)$ exceeds a threshold energy for recognition: $E(\vec t,\vec p) > {E_R}$. T-cell cross-reactivity determines the fraction of peptides recognized by a certain TCR, whereas peptide cross-reactivity determines the fraction of TCRs recognized by a certain peptide.
(B) Distribution of T-cell cross-reactivity and (C) peptide cross-reactivity obtained by examining $10^4$ TCR and $10^4$ peptide sequences for three recognition thresholds, \rev{$E_R = 21 k_B T$} (inverse triangles, red dotted line), \rev{$E_R = 24 k_B T$} (squares, red solid line), and \rev{$E_R = 27 k_B T$} (triangles, red dashed line), where $k_B T$ is the thermal energy at body temperature $T$. For a comparison, the cross-reactivity of random pairs between TCRs and peptides (Random, circles, black solid line) is also plotted. Here the total number of recognition pairs is constrained to be the same with the one for the threshold \rev{$E_R = 24 k_B T$}. The error bars represent the standard errors of the TCR ensembles from nine subjects.}
\label{fig:crossreactivity}
\end{figure}

\subsection{Broad cross-reactivity}
Using the string model with the sequence data ($\vec{t}, \vec{p}$) of TCRs and antigenic peptides, we calculated their binding energy $E(\vec{t}, \vec{p})$, and defined reactive pairs of ($\vec{t}, \vec{p}$) when the pairs have larger binding energies than a recognition threshold, $E(\vec{t}, \vec{p})>E_R$.
Based on the reactive pairs, we defined two kinds of cross-reactivities (Fig. \ref{fig:crossreactivity}A): (i) T-cell cross-reactivity determined the number of peptides recognized by a specific TCR, and (ii) peptide cross-reactivity determined the number of distinct TCRs that recognized a specific peptide. Thus, we measured the fraction of recognized peptides by TCRs and plotted the relative frequency of the cross-reactive fraction (Fig. \ref{fig:crossreactivity}B). Here, we used the fraction of cross-reactive peptides rather than their absolute number because the absolute number depends on the total number of available peptides. Similarly, we measured the fraction of TCRs that recognize the same peptides and plotted their relative frequency (Fig. \ref{fig:crossreactivity}C).
Since we defined the reactivity between a TCR and a peptide as $E(\vec t,\vec p) > {E_R}$, our results depend on the threshold ${E_R}$.
A high/low ${E_R}$ sets a strict/loose threshold for recognizing peptides, which leads to low/high cross-reactivities of TCRs and peptides.
Hereafter, we fixed $E_R = 24 {k_B}T$ of which value was used to explain the thymic selection process of TCRs in a previous study \cite{Kosmrlj2008}.

As a reference for the cross-reactivity distribution, we considered a control scenario that reactive pairs of TCRs and peptides are randomly chosen without considering their binding energy, but the total number of reactive pairs was constrained to be the same with the string model for \rev{$E_R = 24 k_B T$}. Compared with the control cross-reactivity distribution (black lines in Figs.~\ref{fig:crossreactivity}B and \ref{fig:crossreactivity}C), the cross-reactivity distributions of TCRs and peptides were certainly broad regardless of specific choices of $E_R$ (red lines in Figs.~\ref{fig:crossreactivity}B and \ref{fig:crossreactivity}C).
The control scenario predicted that most TCRs reacted with $1 \%$  of total peptides.
Unlike the uniform cross-reactivity, the string model based on the TCR-peptide binding energy predicted a broad cross-reactivity distribution in which some TCRs reacted with $10 \%$ of total peptides whereas some TCRs reacted with only $0.01 \%$ of total peptides.


\begin{figure}[h!]
\centering\includegraphics[width=\linewidth]{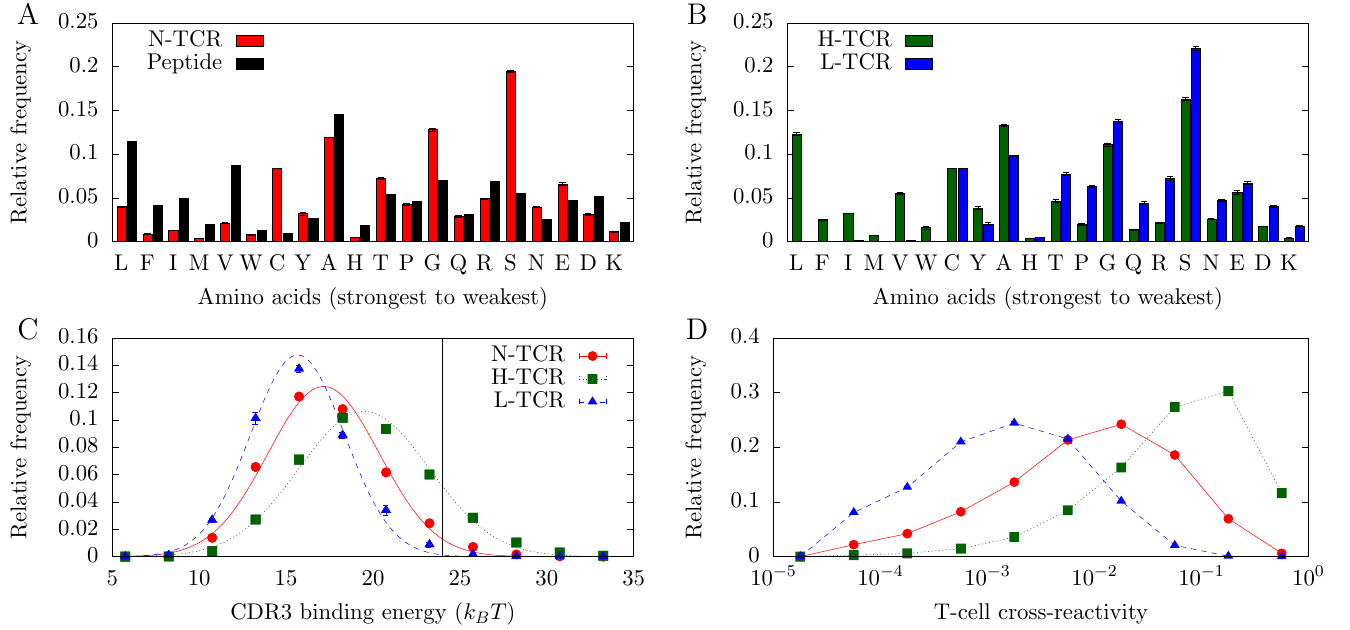}
\caption{Cross-reactivity and amino acid compositions of T-cell receptors. (A) Relative frequencies of twenty amino acids for natural T-cell receptors (N-TCR) and antigenic peptides. Here, the amino acids are sorted from the strongest to the weakest according to its maximum binding energy interacting with all the other amino acids. (B) Relative frequencies of twenty amino acids for top five percentile (H-TCR) and bottom five percentile cross-reactive T-cell receptors (L-TCR). For each amino acid, \rev{most} frequency difference between H-TCR and L-TCR is statistically significant ($P<0.001$) by an unpaired two-tailed Student's $\emph{t}$-test. (C) Binding energy distributions of TCR-peptide pairs for N-TCR (red dots and solid line), H-TCR (green squares and dotted line) and L-TCR (blue triangles and dashed line). Symbols represent simulation results, while lines represent Gaussian distributions obtained by an analytical calculation (see Appendix A). The solid vertical line depicts the recognition threshold of \rev{$E_R = 24 k_B T$}.
(D) T-cell cross-reactivity of three T-cell repertoires. For N-TCR (red circles and solid line), 3,000 natural T-cell receptors were \rev{synthetically generated by using the amino acid compositions in (A)}. For H-TCR (green squares and dotted line) and L-TCR (blue triangles and dashed line), we examine two synthetic T-cell repertoire in which 3,000 TCR sequences are generated by using the amino acid compositions in (B). For (A) and (B), the error bars represent the standard errors of TCR ensembles from nine subjects. For (C) and (D), the error bars represent ten TCR ensembles for synthetic TCRs (N-TCR, H-TCR and L-TCR), which are too small to notice.}
\label{fig:AAcomposition}
\end{figure}

To understand the origin of the variability of the T-cell cross-reactivity, we examined the amino acid compositions of TCRs depending on the degree of cross-reactivity (Figs.~\ref{fig:AAcomposition}A and \ref{fig:AAcomposition}B).
We selected top and bottom 5 percentile cross-reactive TCRs, and denoted them as H-TCR and L-TCR.
Note that H-TCR and L-TCR are subsets of total natural TCRs (N-TCR).
Then, we counted the relative frequency ${f_\alpha }({a_i})$ of the 20 types of amino acids ${a_i},i \in \{ 1,2,...,20\}$ for the three groups of TCRs, $\alpha  \in$ \{N-TCR, H-TCR, L-TCR\}.
As expected, H-TCRs contain more strong amino acids such as leucine, whereas L-TCRs contain more weak amino acids.
Compared with the two groups, N-TCRs contain more moderate amino acids.
We also considered top/bottom 20 percentile cross-reactive TCRs and confirmed that the modification did not qualitatively change the results. This finding is consistent with the previous reports that high cross-reactive TCRs have more amino acids with high hydrophobicity~\cite{Kosmrlj2008,Stadinski2016}, given the correlation between strong amino acids and high hydrophobicity~\cite{Kyte1982}.

We then examined the binding energy distribution between TCRs and peptides for the three groups (Fig.~\ref{fig:AAcomposition}C).
Since we had the amimo acid compositions ($f_\alpha (a_i), g(a_i)$) for TCRs and peptides (Figs.~\ref{fig:AAcomposition}A and \ref{fig:AAcomposition}B), we could estimate the mean and variance of the binding energy (Appendix A), and derive the distribution of binding energy (lines in  Fig.~\ref{fig:AAcomposition}C).
Given the recognition threshold \rev{$E_R = 24 k_B T$}, H-TCR had the most abundant pairs of recognizable TCR-peptide.
To further validate our conclusion, we created synthetic TCR repertoires (denoting again as N-TCR, H-TCR and L-TCR) by assembling amino acids sampled from the distribution ${f_\alpha }({a_i})$ with $\alpha  =$ \{N-TCR, H-TCR, L-TCR\}.
Then, we confirmed that indeed the synthetic H-TCR/L-TCR had more high/low cross-reactive TCRs than N-TCR (Fig.~\ref{fig:AAcomposition}D).
We will use the synthetic repertoires to investigate target searching of T cells.

\begin{figure}[h!]
\centering\includegraphics[width=\linewidth]{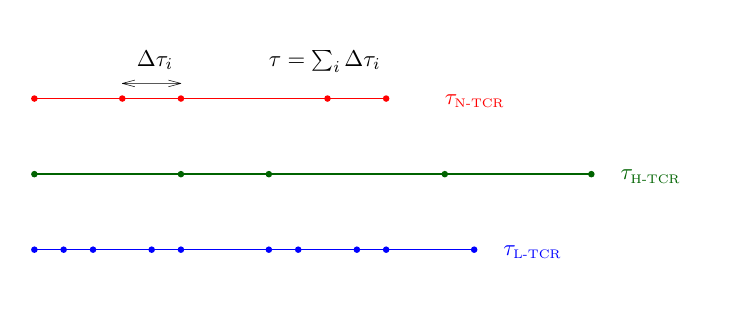}
\caption{Cross-reactivity and target searching of T cells.
Total searching time $\tau $ is a summation of every searching time $\Delta \tau_i = \Delta \tau (E_i)$ that depends on the binding energy $E_i$ between a TCR and a peptide.
Compared with the natural T-cell repertoire (N-TCR), high cross-reactive TCRs (H-TCR) spend longer time to unbind incorrect TCRs from a target peptide, whereas low cross-reactive TCRs (L-TCR) require many trials to find correct TCRs for the target peptide.}
\label{fig:target}
\end{figure}

\subsection{Rapid target searching}
What does the broadness of cross-reactivity implicate? It must be related with many aspects of immunological recognition.
High cross-reactive T cells are effective to cover many antigenic peptides considering the finiteness of the T-cell repertoire in the body.
On the other hand, low cross-reactive T cells can be effective to detect specific antigenic peptides and not to respond to self peptides.
Furthermore, different antigenic peptides can have different degrees of cross-reactivity to T cells depending on their virulence~\cite{Mayer2015}.
In this study, since we found the broad cross-reactivity based on the TCR-peptide binding energy, we explored its functionality in terms of target searching that mostly relies on the TCR-peptide binding and unbinding.

To examine the target searching in a simple setup, we considered a sequential target searching that a certain peptide is searched by a single T cell at a time without time intervals between trials. Then, the total searching time before meeting a right T cell can be defined as
\begin{equation}
\label{eq:totalt}
\tau  = \sum_{i = 1}^{m - 1} \Delta {\tau}(E_i),
\end{equation}
where the $m$-th trial should be the first successful recognition (${E_m} > {E_R}$ and ${E_i} \le {E_R}$ for $i < m $ ).
Each unsuccessful trial spends $\Delta {\tau}(E_i)$ that increases exponentially as the binding energy $E_i$ increases~\cite{Gerland2002}:
\begin{equation}
\label{eq:stept}
\Delta {\tau }(E_i) = \tau_0 \exp (\beta E_i),
\end{equation}
where ${\beta ^{ - 1}} = {k_B}T$ is the inverse thermal energy, and ${\tau _0}$ is a free parameter determining the overall time scale of the searching process.
We set ${\tau _0}$ to make sure that each T cell takes approximately $\left\langle {\Delta {\tau}} \right\rangle  = 5$ minutes on average for the scanning process, considering the observation that immunological synapse formation requires 5 to 30 minutes for antigen recognition from T cells \cite{Huppa2003}.
A more realistic value of ${\tau _0}$ can be smaller, if one considers that the half-lives of the bond between a TCR and agonist pMHC are 1-100 seconds~\cite{Grakoui1999,Holler2003,Krogsgaard2003}.

Given the simple setup, we examined the target searching of the natural T-cell repertoire in which T cells have a broad spectrum of cross-reactivity. We compared the target searching of N-TCR with the target searching of two synthetic repertoires (H-TCR and L-TCR) prepared in the previous section.
H-TCR has high cross-reactivity with strong binding affinity.
Therefore, it is expected that H-TCR does not require many trials to find a right TCR targeting for a certain peptide, but it wastes much time for incorrect TCRs to be unbound from the peptide (Fig.~\ref{fig:target}).
On the other hand, L-TCR does not spend much time for the screening with weak binding affinity, but it requires many trials for the correct targeting.
To evaluate this hypothesis, we estimated mean trial numbers and searching time for the three repertoires.
Since we knew the binding energy distribution of TCR-peptide pairs (Fig.~\ref{fig:AAcomposition}C), we could analytically calculate mean trial number and searching time to sample recognizable TCR-peptide pairs (Appendix B).
However, this feasible analysis calculate the trial number for TCRs to recognize randomly selected peptides at every trial, not a specific peptide.

To evaluate target searching for a specific peptide, therefore, we conducted a target searching simulation given the simple setup.
We first prepared finite numbers of the three T-cell repertoires.
For N-TCR, H-TCR and L-TCR, we generated 3,000 synthetic TCR sequences as described in the previous section.
Given a T-cell repertoire, we introduced a target peptide ${\vec p_i}$ and randomly sampled a TCR ${\vec t_j}$ from the repertoire.
Then, we calculated the binding energy $E({\vec t_j},{\vec p_i})$ and determined whether it exceeded the recognition threshold, $E({\vec t_j},{\vec p_i}) > {E_R}$. Unless recognition was successful, we repeated this process and sampled another TCR until the recognition succeeded at $j = m$.
Finally, ignoring the time interval between trials, we could obtain the total search time in Eq.~(\ref{eq:totalt}) for the target peptide ${\vec p_i}$.

\begin{figure}[h!]
\centering\includegraphics[width=\linewidth]{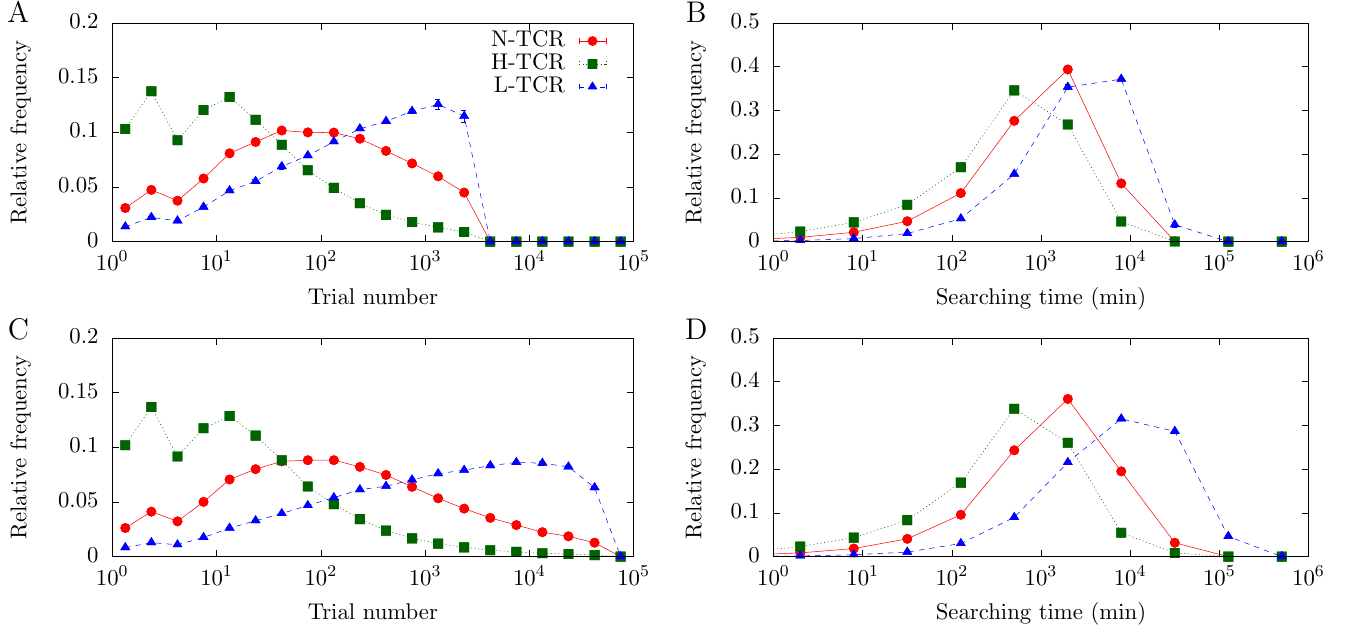}
\caption{Trial number and searching time of T cells for targeting specific peptides. (A) Distribution of trial numbers and (B) searching time of natural T cells (N-TCR, red circles), high cross-reactive T cells (H-TCR, green squares), and low cross-reactive T cells (L-TCR, blue triangles). Each repertoire has 3,000 distinct T cells. (C) Trial number and (D) searching time distributions for larger repertoires with 50,000 distinct T cells. The error bars represent the standard errors of ten ensembles per repertoire, which are too small to notice in some plots.}
\label{fig:targetdist}
\end{figure}

A finite repertoire cannot recognize every peptide in general. L-TCR recognizes fewer peptides than N-TCR, and N-TCR recognizes fewer peptides than H-TCR (Table~\ref{table:recogprob}).
For a fair comparison between the three repertoires, we considered only the specific peptides that were successfully recognized by the three repertoires. Given successful peptide recognition, we counted the trial number $m$ and total searching time $\tau$ and obtained their distributions for the three repertoires (Figs.~\ref{fig:targetdist}A and \ref{fig:targetdist}B).
For targeting specific peptides, N-TCR required \rev{$314 \pm 25$ trials on average, which were about four times of $79 \pm 7$ trials of H-TCR, but N-TCR spent $30 \pm 2$ hrs for target searching, which were only two times longer than $15 \pm 1$ hrs of H-TCR.}
Regarding L-TCR, it \rev{can only target few} peptides (Table~\ref{table:recogprob}), and required extremely many trials and long searching time even for the case of successful targeting (Fig.~\ref{fig:targetdist}A). 
We repeated the same simulation with a larger repertoire of 50,000 TCRs, which achieve more successful recognition (Table~\ref{table:recogprob}).
The larger repertoires allowed more trials and longer searching time that lead to wider distributions for the trial number and searching time (Figs. \ref{fig:targetdist}C and \ref{fig:targetdist}D).
\rev{Then, N-TCR required $1,859 \pm 142$ trials and H-TCR required $261 \pm 21$ trials.
However, unlike the seven-fold difference in the trial number, N-TCR spent $51 \pm 2$ hrs for target searching, which were only less than three times longer than $19 \pm 1$ hrs of H-TCR.}
High cross-reactivity of TCRs is advantageous for quick target searching without extensive screening, but their strong binding affinity is disadvantageous for unbinding from incorrect targeting.
Based on the quantitative simulations with two repertoire sizes, we concluded that N-TCR required many more trials for target searching than H-TCR, but their actual searching time had smaller difference due to the high penalty of unsuccessful trials for H-TCR.


\begin{table}[]
\centering
\begin{tabular}{ccc}
\hline
 &       \multicolumn{2}{c}{Peptide recognition (\%)}                                  \\
\hline
              & TCR pool (3,000)               &  TCR pool (50,000)             \\
\hline
H/N/L-TCR     & $26.74  \pm 3.57$  & $45.03 \pm 2.59$  \\
H/N-TCR       & $56.08 \pm 3.71$ & $49.58 \pm 2.80$ \\
H-TCR         & $15.46 \pm 2.70$ & $4.86 \pm 0.61$ \\
None          & $1.72 \pm 0.47$   & $0.52 \pm 0.11$\\
\hline
\end{tabular}
\caption{Percentage of successful peptide recognition by T cells.
Three T-cell repertoires are considered with natural T-cell receptors (N-TCR), high cross-reactive T-cell receptors (H-TCR), and low cross-reactive T-cell receptors (L-TCR). Each repertoire is considered with ten ensembles of two different pool sizes (3,000 and 50,000).
H/N/L-TCR represents the case that all three repertoires recognize specific peptides,
H/N-TCR represents the case that only H-TCR and N-TCR recognize specific peptides,
H-TCR represents the case that only H-TCR recognizes specific peptides,
and None represents the case that none of the three repertoires recognize specific peptides.}
\label{table:recogprob}
\end{table}

\begin{figure}[h!]
\centering\includegraphics[width=\linewidth]{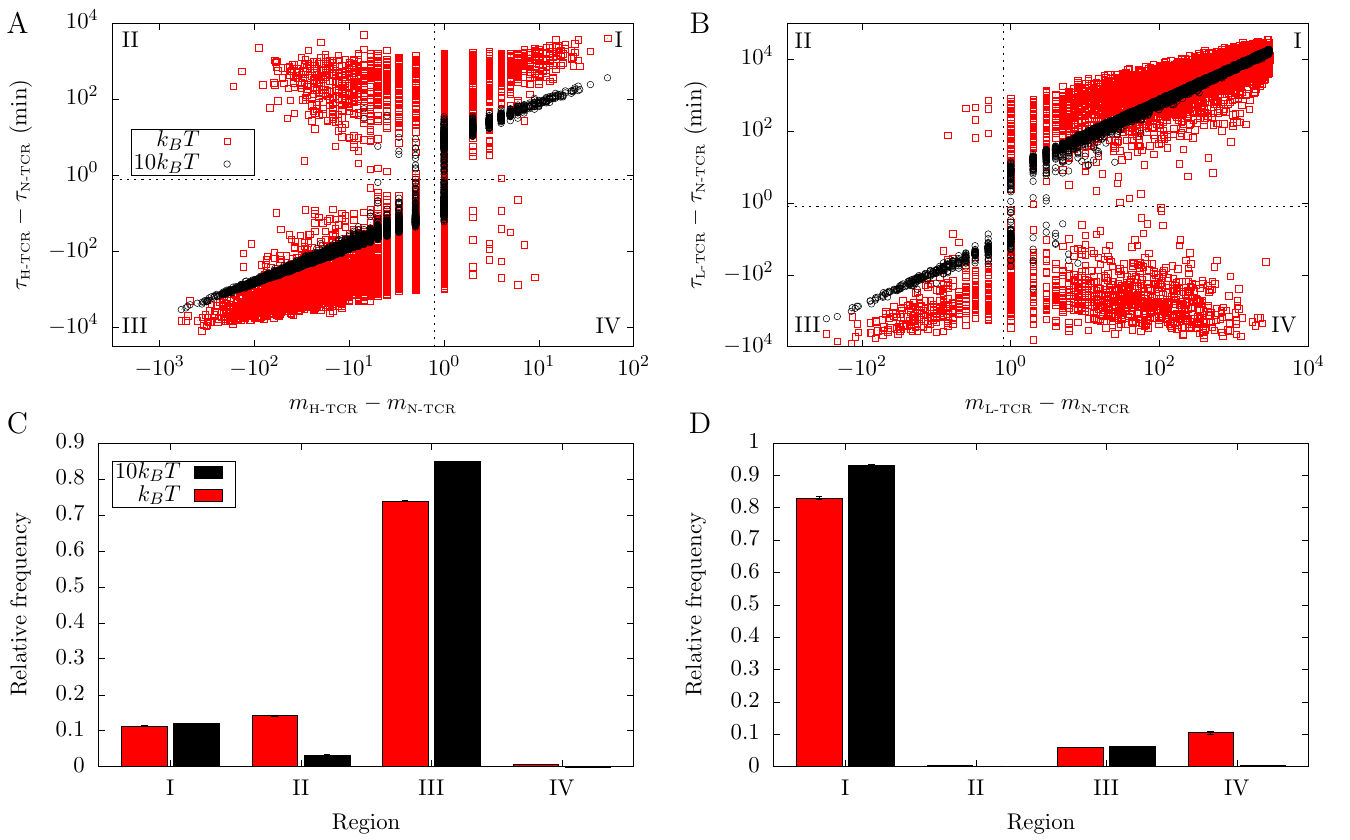}
\caption{Non-trivial relation between trial number and searching time of T cells for targeting specific peptides. The relative trial number and searching time for targeting specific peptides (A) between the natural T-cell repertoire (N-TCR) and high cross-reactive T-cell repertoire (H-TCR), and (B) between the N-TCR and low cross-reactive T-cell repertoire (L-TCR).
Each $i$-th trial spends $\Delta {\tau _i} \propto \exp (\beta {E_i})$ depending on the binding energy ${E_i}$ between a TCR and a peptide.
Note that ${\beta ^{ - 1}} = {k_B}T$ and $10k_B T$ are the thermal energy at body temperature $T$ (red squares) and a higher temperature $10T$ (black circles).
Depending on the relative trial number and searching time, four regions are defined.
(C) The relative frequency of the four regions in (A) for N-TCR versus H-TCR,
and the same plot (D) for N-TCR versus L-TCR. Each repertoire has 3,000 distinct T-cell receptors.
The error bars represent the standard errors of ten ensembles of the three repertoires.}
\label{fig:targetbar}
\end{figure}

To further investigate the penalty of the unbinding time, we observed the relationship between the trial number and searching time for the specific peptides that are successfully recognized by the three repertoires simultaneously (Table~\ref{table:recogprob}).
The trial number was positively correlated with the searching time in general (Figs.~\ref{fig:targetbar}A and \ref{fig:targetbar}B).
However, the penalty of the unbinding time occasionally reversed the positive correlation.
For certain peptides, N-TCR required more trials to recognize them than H-TCR, but N-TCR could search them faster than H-TCR ($m_{\rm{H-TCR}} < m_{\rm{N-TCR}}, \tau_{\rm{H-TCR}} > \tau_{\rm{N-TCR}}$).
\rev{The negative correlation between $m$ and $\tau$ happened for $14.2 \pm 0.7\% $ between H-TCR and N-TCR, and $10.6 \pm 1.6\% $ between N-TCR and L-TCR (II and IV regions in Figs.~\ref{fig:targetbar}C and \ref{fig:targetbar}D).}
It is noteworthy that the body temperature is crucial for this conclusion.
If the temperature increased 10-fold (${\beta ^{ - 1}} = 10{k_B}T$) in Eq.~(\ref{eq:stept}), the penalty of unbinding diminished, and $m$ and $\tau$ became strongly correlated (black symbols in Figs.~\ref{fig:targetbar}C and \ref{fig:targetbar}D). H-TCRs had the fastest target searching with the fewest trials. In contrast, if the temperature decreased 10-fold (${\beta ^{ - 1}} = 0.1{k_B}T$), the unbinding penalty became substantial, and H-TCRs had the slowest target searching (data not shown).

Thus far, we ignored the recognition process at the $m$-th event of correct targeting. However, if we considered the synapse dissolution process is necessary to induce immune responses including T-cell proliferation~\cite{Huppa2003}, we may need to include the $m$-th binding event into the searching time as $\tau'  = \sum_{i = 1}^{m} \Delta \tau (E_i)$.
If the $m$-th targeting event was included, the longer unbinding time of H-TCR became more penalized, and the mean searching time of N-TCR became shorter than those of H-TCR and L-TCR: $\langle \tau'_{\rm{N-TCR}} \rangle < \langle \tau'_{\rm{L-TCR}} \rangle < \langle \tau'_{\rm{H-TCR}} \rangle$.

\section{Summary}
Understanding the recognition of specific peptides by individual T cells is a central problem in the adaptive immunity.
Although structure-based approaches solved various questions, they are limited to investigate general principles of the immunological recognition from a large-scale analysis for the TCR-peptide recognition.
However, recent high-throughput sequencing technologies provide digital information regarding TCRs~\cite{Warren2011,Murugan2012,Zvyagin2014}.
Furthermore, theoretical immunologists have developed a string model that estimates the binding affinity between a TCR sequence and a peptide sequence~\cite{Kosmrlj2008, Kosmrlj2009}.
In this study, we adopted the string model to examine the cross-reactivity of human T cells by using published TCR sequences~\cite{Murugan2012} and antigenic peptide sequences~\cite{Vita2015}.

By examining the large-scale recognition between $10^4$ TCRs and $10^4$ antigenic peptides, we found that T cells have a broad spectrum of cross-reactivity.
Some T cells are reactive to $1,000$ peptides, but some T cells are reactive only to one or two peptides.
\rev{This result supports the huge variability of the cross-reactivity reported in recent experiments.}
Certain T cells can respond to $10^6$ different peptides, whereas other T cells have an extremely low cross-reactivity to different peptides despite their sequence similarity~\cite{Petrova2012}. 
In addition to the T-cell cross-reactivity, populations of naive T cells specific for some dominant peptides were $10$ times larger than those specific for some less dominant peptides~\cite{Jenkins2012,Maillere2013}.
We emphasize that the current experiments on the T-cell and peptide cross-reactivity are limited to examine a few T cells with a large pool of peptides, or a few peptides with a large pool of T cells.
However, the string model allowed to computationally estimate a full spectrum of the cross-reactivity from $10^4$ TCRs and $10^4$ peptides.

Here, we confirmed that high/low cross-reactive TCRs had higher/lower frequencies of strongly interacting amino acids as reported in previous studies~\cite{Kosmrlj2008,Stadinski2016}.
The degree of cross-reactivity should have a positive correlation with the (un)binding affinity of TCRs.
Therefore, we investigated how the broad spectrum of cross-reactivity affected the target searching of T cells.
We expected that high cross-reactive T cells do not require many trials for searching correct targets, but they spend long time to unbind from incorrect targets. In contrast, low cross-reactive T cells require many trials for screening correct targets, but they spend short time to ignore incorrect targets.
We examined the target searching of the natural T cells, having the broad cross-reactivity, by comparing it with two synthetic T-cell repertoires, having high and low cross-reactivity. Then, we evaluated the above hypothesis, and showed that the disadvantages of the high and low cross-reactive T cells were highlighted at a physiological temperature. The broad cross-reactivity could be a trade-off for balancing the rapid screening and unbinding penalty.

Given the plethora of sequence data, we attempted to extract a general principle in the immunological recognition by using a string model. The simple model allowed a large-scale analysis, but its limitations should be noted for more realistic understanding of the immunity.
The model assumed the linear interaction between CDR3 and pMHC considering their short interface, but their real interaction should be determined by their three-dimensional structures and dynamics.
For the analysis of target searching, the estimated searching time should be cautiously considered as absolute time, because simultaneous target searching of multiple T cells and the time intervals between searching events were ignored for simplicity. 
\rev{Since we considered peripheral T cells filtered out from the negative selection in the thymus, we assumed that their interaction to self-peptides was negligible in target searching.}
Furthermore, we did not consider the details of MHC or its preference for specific peptides (MHC restriction)~\cite{Calis2010}, the tolerance on self-peptides~\cite{Deng2007}, and the suppression of regulatory T cells~\cite{Sojka2008}, which can be essential ingredients for understanding the cross-reactivity and target searching of T cells.

\appendix

\setcounter{table}{0}
\renewcommand\thetable{\Alph{section}}
\section{Binding energy and recognition probability of a TCR-peptide pair}
As previously reported \cite{Kosmrlj2009,Derrida1980,Derrida1981}, we approximated the distribution of a TCR-peptide binding energy $E$ as a Gaussian distribution:
\begin{equation}
\label{eq:bindingEdist}
{P_\alpha }(E) = \frac{1}{{\sqrt {2\pi \sigma _\alpha ^2} }}\exp \left[ { - \frac{{{{(E - {\mu _\alpha })}^2}}}{{2\sigma _\alpha ^2}}} \right]
\end{equation}
with mean ${\mu _\alpha }$ and variance $\sigma _\alpha ^2$ for different T-cell repertories $\alpha$.
Given the amino acid compositions $f_{\alpha}(a)$ for each T-cell repertoire and $g(a)$ for antigenic peptides, we can estimate the mean and variance of the binding energy between a TCR sequence and a peptide sequence with an effective length {\color{blue}$L_{\text{eff}} = 6$:}
\rev{
\begin{align}
\label{eq:Eavg}
\langle E \rangle _\alpha &= L_{\text{eff}}  \sum_{i,j=1}^{20} {{f_\alpha }({a_i})g({a_j})J({a_i},{a_j})}, \\
\label{eq:Eerror}
\langle E^2 \rangle _\alpha  &=  L_{\text{eff}}  \sum_{i,j=1}^{20} {{f_\alpha }({a_i})g({a_j}){J^2}({a_i},{a_j})} \nonumber \\
& + ({L_{\text{eff}}^2} - L_{\text{eff}})\sum_{i \ne k,j \ne l;i,j,k,l = 1}^{20} {{f_\alpha }({a_i})g({a_j}){f_\alpha }({a_k})g({a_l})J({a_i},{a_j})} J({a_k},{a_l}).
\end{align}
}
The analytical estimation is consistent with the mean and variance of the simulated binding energy distributions (Table~\ref{table:bindingapprox}). In addition, the good agreement for $\alpha$=N-TCR suggests that the site-site correlation was negligible in the amino acid sequence of the natural TCRs.
It is expected that high cross-reactive TCRs (H-TCR) have higher binding energies with peptides than low cross-reactive TCRs (L-TCR).
The stronger binding affinity is the source of high cross-reactivity. To further quantify this relation, we defined a probability ${q_\alpha }$ that a TCR-peptide pair had stronger binding than the recognition threshold ${E_R}$:
\begin{equation}
\label{eq:recogprob}
{q_\alpha } = \int_{{E_R}}^\infty  {{P_\alpha }(E)} dE.
\end{equation}
The theoretical estimation of ${q_\alpha }$ could reasonably predict the success probability of TCR-peptide recognition in the simulation (Table~\ref{table:bindingapprox}). 

\begin{table}[]
\centering
\begin{tabular}{cccc}
\hline
     $\alpha$                &${\mu _\alpha}(k_BT)$         & ${\sigma_\alpha^2}(k_B^2 T^2)$ & ${q_\alpha} (\%)$         \\
\hline
H-TCR                  & $19.63 \pm 0.04$              & $14.39 \pm 0.07$              & $12.90 \pm 0.15$   \\
                       & ($19.49 \pm 0.03$)            & ($14.07 \pm 0.05$)            & ($11.46 \pm 0.21$) \\
\hline
N-TCR                  & $17.30 \pm 0.03$              & $10.55 \pm 0.07$              & $2.96 \pm 0.08$   \\
                       & ($17.16 \pm 0.03$)            & ($10.26 \pm 0.04$)            & ($1.64 \pm 0.05$)               \\
\hline
L-TCR                  & $15.79 \pm 0.02$              & $7.57 \pm 0.03$              & $0.42 \pm 0.01$               \\
                       & ($15.66 \pm 0.01$)            & ($7.30 \pm 0.02$)            & ($0.1$)            \\
\hline
\end{tabular}
\caption{Mean and variance of the binding energy between T-cell receptors and peptides, and their recognition probability.
We obtained the mean $\mu_\alpha$ and variance $\sigma_\alpha^2$ of the TCR-peptide binding energy from the numerical simulations and analytical estimations (parentheses) for three T-cell repertoires $\alpha$: natural T-cell receptors (N-TCR), high cross-reactive T-cell receptors (H-TCR), and low cross-reactive T-cell receptors (L-TCR). The probability ${q_\alpha }$ of successful recognition was also estimated for the event when the binding energy exceeded the threshold energy for the peptide recognition using the two methods. The errors were estimated from ten ensembles of the repertoires. Too small errors for $q_\alpha$ are not shown.}
\label{table:bindingapprox}
\end{table}

\section{Trial number and searching time of T cells for targeting peptides}
We estimated the mean searching time for TCRs to recognize anonymous peptides among a set of peptides.
First, we sampled a binding energy ${E_i}$ from the binding energy distribution (Fig.~\ref{fig:AAcomposition}C), until the sampled ${E_i}$ exceeded a recognition threshold ${E_R}$.  To estimate the total searching time $\tau $, we consider the probability that successful recognition occurs at the $m$-th trial:
\begin{equation}
\label{eq:recogmth}
{Q_\alpha }(m) = {(1 - {q_\alpha })^{m - 1}}{q_\alpha },
\end{equation}
that assumes $m$ sequences of independent events with $(m-1)$ failures and one final success with success probability ${q_\alpha }$ in Eq.~(\ref{eq:recogprob}). Here, the mean searching time per trial can be also estimated from the binding energy distribution ${P_\alpha }(E)$:
\begin{equation}
\label{eq:timepertrial}
{\left\langle {\Delta \tau } \right\rangle _\alpha } = \int_{ - \infty }^{{E_R}} {{\tau _0}} {e^{\beta E}}{P_\alpha }(E)dE.
\end{equation}
Given the mean searching time per trial, the total mean searching time for $(m-1)$ trials is defined as
\begin{equation}
\label{eq:timetotal}
{\tau _\alpha }(m) \approx (m - 1){\left\langle {\Delta \tau } \right\rangle _\alpha }.
\end{equation}
Then, we can estimate the mean total searching time,
\begin{eqnarray}
\label{eq:avgtimetotal}
\langle \tau \rangle &=& \sum_{m = 1}^\infty  \tau _\alpha(m) Q_\alpha (m) \nonumber \\
&\approx&  \langle \Delta \tau \rangle_\alpha \sum_{m = 1}^\infty  (m - 1)(1 - q_\alpha )^{m - 1} q_\alpha \nonumber \\
&\approx& \langle \Delta \tau \rangle_\alpha \frac{1 -  q_\alpha}{q_\alpha},
\end{eqnarray}
where the last approximation holds for $0 < {q_\alpha } < 1$.
The final equality in Eq.~(\ref{eq:avgtimetotal}) can define the mean trial number,
\begin{equation}
\label{eq:avgtrial}
{\left\langle {m - 1} \right\rangle _\alpha } = \frac{{1 - {q_\alpha }}}{{{q_\alpha }}}.
\end{equation}
Using the analytical formulation, we could estimate the mean trial number $\langle {m - 1} \rangle _\alpha$ and mean searching time $\langle \tau \rangle_\alpha$ for the three T-cell repertoires (Table~\ref{table:targetapprox}). 
Here, the smaller difference in the total searching time compared with trial number was due to the penalty of strong unbinding.

\begin{table}[]
\centering
\begin{tabular}{ccc}
\hline
$\alpha$ &${\left\langle {m - 1} \right\rangle _\alpha }$ & ${\left\langle \tau  \right\rangle _\alpha }$ (min)                                 \\
\hline
H-TCR                                           & $8$                              & $(3.69 \pm 0.05) \times {10^2}$ \\
N-TCR                                           & $60 \pm 2$                       & $(1.10 \pm 0.01) \times {10^3}$ \\
L-TCR                                           & $987 \pm 31$                     & $(4.18 \pm 0.06) \times {10^3}$ \\
\hline
\end{tabular}
\caption{Mean trial number and searching time of T cells for targeting peptides.
The mean trial number $\langle m - 1 \rangle_\alpha$ and searching time $\langle \tau \rangle_\alpha$ were estimated analytically from the TCR-peptide binding energy distributions for three T-cell repertoires: natural T-cell receptors (N-TCR), high cross-reactive T-cell receptors (H-TCR), and low cross-reactive T-cell receptors (L-TCR). The errors were estimated from ten ensembles of the repertoires.}
\label{table:targetapprox}
\end{table}

\section*{Acknoledgement}
We thank M. Kardar and A.K. Chakraborty for critical discussions and hospitality during J.X.'s visit at the Massachusetts Institute of Technology, where a part of this study was performed. And we also appreciate for the valuable comments and suggestions from reviewers to improve the quality of this paper. This study was supported by the Basic Science Research Program of the National Research Foundation of Korea (NRF), the Ministry of Education (2016R1D1A1B03932264), the Max Planck Society, and the Korea Ministry of Education, Science and Technology, Gyeongsangbuk-Do and Pohang City (J.J.).

\section*{References}

\bibliography{cross-reactivity}

\end{document}